\documentclass[conference]{IEEEtran}
\IEEEoverridecommandlockouts
\usepackage{cite}
\usepackage{amsmath,amssymb,amsfonts}
\usepackage{algorithmic}
\usepackage{booktabs}
\usepackage{mathtools}
\usepackage{amsmath}
\usepackage{graphicx}
\usepackage{textcomp}
\usepackage{caption}
\usepackage{xcolor}
\def\BibTeX{{\rm B\kern-.05em{\sc i\kern-.025em b}\kern-.08em
    T\kern-.1667em\lower.7ex\hbox{E}\kern-.125emX}}
\begin{document}

\title{Fully Convolutional Neural Networks for Automotive Radar Interference Mitigation
\thanks{The authors acknowledge all national funding authorities and the ECSEL Joint Undertaking, which funded the
PRYSTINE project under the grant agreement 783190. The real-world radar signals used for experimental validation were provided by NXP under the PRYSTINE cooperation agreement. Additionally, this work was co-funded through the Operational Program Competitiveness 2014-2020, Axis 1, contract no. 3/1.1.3H/24.04.2019, MySMIS ID: 121784.}
}

\author{\IEEEauthorblockN{Nicolae-C\u{a}t\u{a}lin Ristea}
\IEEEauthorblockA{
\textit{University Politehnica of Bucharest}\\
Bucharest, Romania \\
r.catalin196@yahoo.ro}
\and
\IEEEauthorblockN{Andrei Anghel}
\IEEEauthorblockA{
\textit{University Politehnica of Bucharest}\\
Bucharest, Romania \\
andrei.anghel@munde.pub.ro}
\and
\IEEEauthorblockN{Radu Tudor Ionescu}
\IEEEauthorblockA{
\textit{University of Bucharest}\\
Bucharest, Romania\\
raducu.ionescu@gmail.com}
}

\maketitle

\begin{abstract}
The interest of the automotive industry has progressively focused on subjects related to driver assistance systems as well as autonomous cars. Cars combine a variety of sensors to perceive their surroundings robustly. Among them, radar sensors are indispensable because of their independence of lighting conditions and the possibility to directly measure velocity. However, radar interference is an issue that becomes prevalent with the increasing amount of radar systems in automotive scenarios. In this paper, we address this issue for frequency modulated continuous wave (FMCW) radars with fully convolutional neural networks (FCNs), a state-of-the-art deep learning technique. We propose two FCNs that take spectrograms of the beat signals as input, and provide the corresponding clean range profiles as output. We propose two architectures for interference mitigation which outperform the classical zeroing technique. Moreover, considering the lack of databases for this task, we release as open source a large scale data set that closely replicates real world automotive scenarios for single-interference cases, allowing others to objectively compare their future work in this domain. The data set is available for download at: http://github.com/ristea/arim.
\end{abstract}


\setlength{\abovedisplayskip}{2pt}
\setlength{\belowdisplayskip}{2pt}

\section{Introduction}

Nowadays, automotive radar sensors are essential elements of driving assistance systems and autonomous driving applications. Their main goal is to estimate the distance and velocity of objects on the road. However, the technical requirements increased steadily from simple detection to braking functions and smart environment perception tasks \cite{Shibao-VTC-2019} for self-driving cars. The most common radar senors used in the automotive industry are frequency modulated continuous wave (FMCW) / chirp sequence (CS) radars, which transmit sequences of linear chirp signals. Furthermore, the amount of automotive radar systems on the streets is regularly increasing \cite{kunert2012eu}, and the high number of automotive radar systems on roads leads to a higher probability of interference between radar sensors. 

\begin{figure}[t]
\centering
  \includegraphics[width=0.88\linewidth]{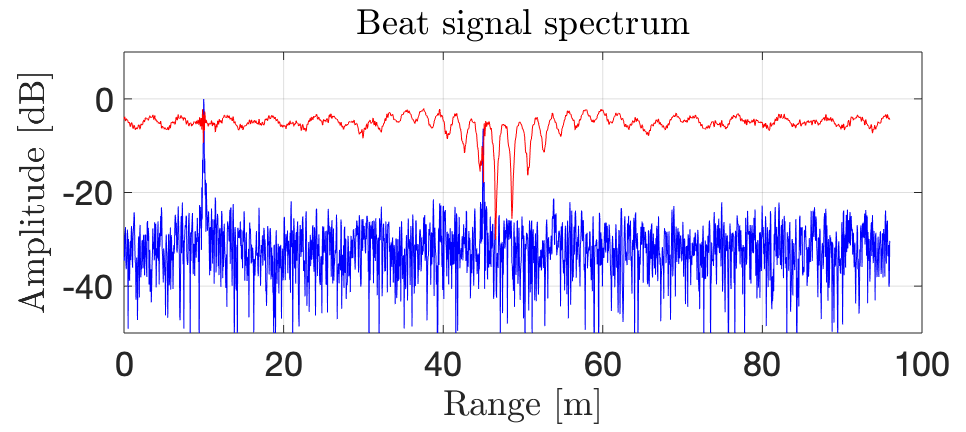}
  \vspace{-0.1cm}
  \caption{Range profile magnitude of an FMCW radar sensor. The useful profile is shown in blue, while the profile with interference is shown in red.}
  \label{fig_1}
  \vspace{-0.3cm}
\end{figure}

Radio frequency interference (RFI) is a relevant issue for radar sensors as it could increase the effective noise floor, reduce sensitivity or create false detections \cite{brooker2007mutual}. We illustrate the negative effects of RFI in Figure \ref{fig_1}, where the noise floor has risen with approximately $15\;dB$ and the targets become almost undetectable. Therefore, the interference mitigation task is a crucial part of current and future radar sensors used in a traffic safety context.

In this paper, we propose a novel approach for radar interference mitigation that is based on fully convolutional neural networks (FCNs) \cite{Long-CVPR-2015}. This architecture is able to learn different types of patterns with a relatively small amount of learnable parameters, a fact that recommends them for real-time applications, such as autonomous driving. Our proposed architecture is fed with spectrograms of beat signals with noise and interference. The output of our neural network is the range profile magnitude with mitigated noise and interference. Additionally, because of the lack of public databases, we propose a novel synthetic database, with signals affected by an interference source, for the community to have a common ground for the evaluation and comparison of future methods.

In summary, our contribution is twofold:
\begin{itemize}
\item We propose a novel approach for interference mitigation based on FCNs, transforming spectrograms into range profiles.
\item We propose a radar interference data set with a wide and realistic range of signal parameter variations to be used as benchmark in future research.
\end{itemize}

\section{Related Work}

\noindent
{\bf Classical methods.}
State-of-the-art interference detection (mitigation) methods are typically classified according to the domain in which the interference is mitigated \cite{mosarim2010d15, uysal2019sync, kim2018peer, xu2018orthogonal_noise, bechter2017dbfmimo, laghezza2019nxp}: polarization, time, frequency, code and space. Polarization-based methods assume the use of cross-polarized antennas between the two interfering radars and the mitigation margin is around 20 dB, but reflections on the ground or other surrounding targets can severely reduce this margin. Time domain methods include the following approaches: use of a low transmit duty cycle (to reduce the probability of hitting other receivers), use of a short receive window (to reduce the probability of being hit by an interferer), or employing a variable pause between chirps or a variable chirp slope (to avoid periodic interferences). Frequency domain methods imply a division of the authorized operating bandwidth into several sub-bands, such that nearby radars operate in different sub-bands. RFI mitigation in the coding domain involves the modulation of the radar waveforms with a device-specific code (to minimize cross-talk between radars, the codes of different devices should be orthogonal), whereas in the case of space domain techniques, the antenna radiation pattern is adaptively configured to avoid interfering signals. 

A particular class of methods are the strategic RFI mitigation techniques \cite{mosarim2010d15}, which involve additional hardware and/or software, yet rely on some of the basic techniques. The classical strategic approaches are: ``communicate and avoid'' (requires inter-vehicle communication to avoid simultaneous transmission), ``detect and avoid'' (e.g., detects the interference in a sub-band and changes the operating sub-band of the radar), ``detect and repair'' (after detection, the measurement with interference is reconstructed), ``detect and omit'' (after detection, the measurements affected by interference are removed) and ``listen before talk'' (the radar transmits only when no other device is detected). 


Different from all these methods, which rely on algorithms handcrafted by researchers, we propose an approach based on end-to-end learning from data. In order to obtain our approach, we constructed a realistic data set to learn deep neural networks.

\noindent
{\bf Deep learning methods.}
Deep learning techniques have been applied in a wide range of tasks with remarkable results \cite{soviany2018optimizing,georgescu2020convolutional}. One such task is image denoising, where deep learning achieved state-of-the-art results \cite{bricman2018coconet}, outperforming classical approaches such as median or bilateral filtering. By transforming the radio signal into a spectrogram, the task of interference mitigation becomes similar to a task of image denoising. In this context, we propose to apply fully convolutional networks, a deep learning technique, to transform a noisy spectrogram into a clean range profile of a radar sensor.
To our knowledge, there are only a handful of related works \cite{rock2019complex,fan2019interference,mun2018deep} that employ deep learning models for radar interference mitigation, but they have different architectures that are based on input and output pairs from the same domain, e.g. both their input and output are spectrograms \cite{rock2019complex}.
In \cite{rock2019complex}, the authors proposed a convolutional neural network (CNN) to address RFI, aiming to reduce the noise floor while preserving the signal components of detected targets. The CNN architecture can be trained using either range processed data or range-Doppler (RD) spectra as inputs. The authors reported promising results, but they still have concerns regarding the generalization capacity on real data. 
Another approach that relies on CNNs is proposed in \cite{Fuchs2020}. The authors employ an autoencoder based on the U-Net architecture \cite{ronneberger2015u}, which performs interference mitigation as a denoising task directly on the range-Doppler spectrum. They surpass classical approaches, but the phase information cannot be fully preserved. Similarly, in \cite{fan2019interference}, the network architecture is build upon CNNs, but residual connections, inspired from \cite{he2016deep}, were added.
A different method is proposed in \cite{mun2018deep}, which is based on applying a recurrent neural network model with Gated Recurrent Units \cite{chung2014empirical} on the time domain signal to mitigate the interference. The authors reported better performance compared to existing signal processing methods and lower processing times.

\section{Method}

\begin{figure*}[!th]
\centering
  \includegraphics[width=0.89\linewidth]{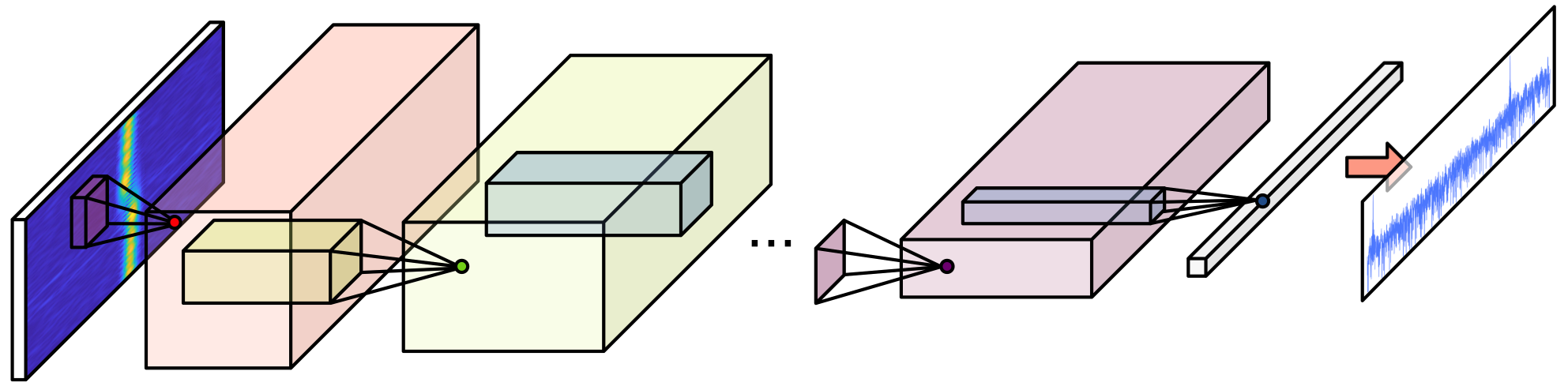}
  \vspace{-0.15cm}
  \caption{The general architecture of our FCN models. The input spectrogram is processed through a series of conv blocks (composed of conv and pooling layers) until the vertical dimension is reduced to 1, while preserving the horizontal dimension. The output is a range profile without the interference removed by the FCN.}
  \label{fig_2}
  \vspace{-0.3cm}
\end{figure*}

\noindent
{\bf Radar signal model.}
In FMCW radar solutions, the transmitted signal $s_{TX}(t)$ is a chirp sequence, whose frequency usually follows a sawtooth pattern. The analytical signal $s_{TX}(t)$ in a sweep interval is defined as follows:
\begin{equation}
\label{x_sign}
s_{TX}(t) =
\left\{
	\begin{array}{ll}
		e^{j 2\pi (f_{0}t + \frac{\alpha t^2 }{2})},  & \mbox{if}\;\; 0 \leq t \leq T_{c} \\
		0, &  \mbox{otherwise}
	\end{array}
\right.,
\end{equation}
where $t$ is the time domain variable, $f_{0}$ is the frequency at the initial moment $t=0$, $T_{c}$ is the frequency variation time interval (sweep time) and $\alpha$ is called the chirp rate calculated as $\alpha = \frac{B}{T_{c}}$, where $B$ is the bandwidth.

The receive antenna collects the reflected signal $s_{RX}(t)$, which, for a single target, is defined as follows:
\begin{equation}
\label{rx_sign}
    s_{RX}(t) =  \underline{A} \cdot s_{TX}(t-\tau) ,
\end{equation}
where $\tau$ is the propagation delay and $\underline{A} = A \cdot e^{j\phi}$ is the complex amplitude of the received signal.

The reflected signal is further mixed with the transmitted signal and low-pass filtered, resulting in the beat signal $s_{b}(t)$, which is analytically computed as: 
\begin{equation}
\label{mix_sign}
    s_{b}(t) =  s_{TX}(t) \cdot s^{\ast}_{RX}(t),
\end{equation}
where $s^{\ast}_{RX}(t)$ represents the complex conjugate of $s_{RX}(t)$.

\noindent
{\bf Interference signal model.}
In the presence of mutual interference, the radar transmits a signal which is reflected by a target and the receiving antenna collects a mix from two signals, the reflected signal and the interference signal, respectively. Consequently, the received signal is defined as follows:
\begin{equation}
\label{int_sign}
s_{RX}(t) = \displaystyle\sum_{j=0}^{N_{t}} \underline{A}_{j} \cdot s_{TX}(t-\tau_{j}) + \displaystyle\sum_{k=0}^{N_{int}} s_{RFI,k}(t),
\end{equation}
where $N_{t}$ is the number of targets, $N_{int}$ is the number of interferers and $\underline{A}$, $\tau$ are the corresponding parameters from Equation~\eqref{rx_sign}. An uncorrelated interfering signal $s_{RFI,k}(t)$ (with a different modulation rate than the one of the transmitted chirp) translates, after mixing with the transmitted signal, into a chirp signal whose bandwidth is limited by the anti-aliasing filter. Therefore, an uncorrelated interference appears as a highly non-stationary component on the spectrogram, which is spread across multiple frequency bins. In the following, we limit the number of interferers to $N_{int}$ = 1 and we consider only uncorrelated interfering sources. 

\noindent
{\bf Data preprocessing.}
Fully convolutional neural networks attain state-of-the-art results in computer vision, the convolutional operation being specifically designed for images. In order to apply FCNs to our task, we first need to transform the time domain signals into images. One of the most common approaches to obtain an image representation of a time domain signal is by computing a spectrogram using the discrete Short-Time Fourier Transform (STFT), as shown in the following equation:
\begin{equation}
\label{stft}
    STFT\{x[n]\}(m, k) = \sum_{n=-\infty}^{\infty} x[n] \cdot w[n-m R] e^{-j \frac{2 \pi}{N_x}k n},
\end{equation}
where $x[n]$ is the discrete input signal, $w[n]$ is a window function, $N_x$ is STFT length and $R$ is the hop/step size \cite{allen1977stft}. There are a plethora of window functions proposed in literature, such as \textit{hann}, \textit{blackman} and others. We chose to perform the STFT with \textit{hamming} window.


Moreover, several time-frequency representations have been developed over time. Wavelet analysis, through the \emph{continuous wavelet transform} with different base functions, can also be used for this purpose. Nevertheless, in this study, we restrict ourselves to the spectrogram, and leave other time-frequency representations for future investigations.

Our goal is to obtain clean range profiles from beat signal spectrograms affected by noise and uncorrelated interference. We design our FCNs to provide the clean range profiles as output (during training, the FCNs have to learn to reproduce the ground-truth clean range profiles). For this reason, we perform a Fast Fourier Transform (FFT) of our time domain labels (to obtain the ground-truth clean range profiles) and train our networks to map the STFT input to the FFT output. The FFT is computed as follows:
\begin{equation}
\label{fft}
    \textrm{FFT}\{x[n]\}(k) = \sum_{n=0}^{N_x-1} x[n] \cdot e^{-j \frac{2 \pi}{N_x}k n},
\end{equation}
\noindent
where $N_x$ is the number of FFT points (same value as the number of STFT range bins).

\noindent
{\bf Network architectures.}
We consider a beat signal with 1024 samples, and we design two FCN models for interference mitigation from single-channel spectrograms. The first FCN takes as input a spectrogram of $154 \times 2048$ components, while the second FCN takes as input a spectrogram of $1024 \times 2048$ components. The horizontal axis dimension ($2048$) corresponds to the number of FFT points in the range profile, while the vertical axis dimension ($154$ or $1024$) is influenced by the computing step size of STFT. The dimensions of $154$ or $1024$ correspond to steps $R\!=\!6$ and $R\!=\!1$, respectively. We design each FCN architecture to progressively reduce the dimension on the vertical axis to one component, while preserving the horizontal axis dimension. Therefore, both FCN models produce outputs of $1 \times 2048$ components that are interpreted as range profiles. In order to reach the same output size from different input sizes, the two FCN models have different depths. In Figure \ref{fig_2}, we illustrate the generic architecture of both networks, which are exclusively composed of convolutional (conv) blocks.

To reduce the spectrogram of $154 \times 2048$ to a range profile of $1 \times 2048$ components, we propose a shallower FCN of 15 layers organized into 4 conv blocks. Each of the first 3 blocks are composed of 3 conv layers followed by a max-pooling layer, while the last block has 3 conv layers (without pooling). Each conv layer in the first block is formed of 8 filters. The number of conv filters doubles in each subsequent block. Based on this rule, the first two conv layers from the fourth block are formed of 64 filters. The last conv layer needs to squeeze the number of channels to one, hence it can have only one filter.

To reduce the spectrogram of $1024 \times 2048$ to a range profile of $1 \times 2048$ components, we propose a deeper FCN of 21 layers organized into 7 conv blocks. Each of the first 6 blocks are composed of 2 conv layers followed by a max-pooling layer, while the last block has 3 conv layers (without pooling). Each conv layer in the first block is formed of 8 filters. The number of conv filters doubles in each subsequent block, except for the second and the fourth blocks, which keep the number of filters from the previous blocks. As for the shallow FCN, the last conv layer has only one filter.

For both architectures, the conv filters with $5\times5$ spatial support are applied at stride 1. Zero padding is added to preserve the horizontal dimension of the activation maps. Except for the very last conv layer, all convolutional layers are followed by ReLU activations. The pooling filters are of size $2\times1$, reducing the size of the activation maps on the vertical axis only. Zero padding for the max-pooling layers is added only when we need to make sure that the input activation maps have an even size. Both networks are trained using the Adam optimizer~\cite{Kingma-ICLR-2015}, using the mean squared error as loss function.

\section{Data Set}

\begin{table}
\centering
\caption{Minimum and maximum values for each parameter in our joint uniform distribution used for generating the samples in our database.}
\label{tab1} 
 \begin{tabular}{|l|ccc|}
 \hline
 Parameter & Minimum & Maximum & Step \\ 
 \hline\hline
 SNR [dB] & 5 & 40 & 5 \\ 
 \hline
 SIR [dB] & -5 & 40 & 5 \\
 \hline
 Relative interference signal slope & 0 & 1.5 & 0.1 \\
 \hline
 Number of targets & 1 & 4 & - \\
 \hline
 Target amplitude & 0.01 & 1 & - \\
 \hline
 Target distance [m] & 2 & 95 & - \\ 
 \hline
 Target phase [rad] & $-\pi$ & $\pi$ & - \\ 
 \hline
\end{tabular}
\end{table} 

\begin{table}
\centering
\caption{Fixed parameters for simulating a realistic radar sensor.}
\label{tab2} 
 \begin{tabular}{|clc|}
 \hline
 Parameter & Description & Value\\
 \hline\hline
 B & Bandwidth & 1.6 GHz\\ 
 \hline
 $T_{r}$ & Time of chirp & 25.6 $\mu s$\\ 
 \hline
 $f_{s}$ & Sampling frequency & 40 MHz\\ 
 \hline
 $f_{0}$ & Radar central frequency & 78 GHz\\ 
 \hline
\end{tabular}
\vspace{-0.3cm}
\end{table}

It is well known that a large database is a key factor in the training process of deep neural networks with high generalization capacity. To the best of our knowledge, there are no public in-the-wild (or generated) databases for the interference mitigation task, mainly because of the difficulty imposed by the process of data acquisition and data labeling. In this context, it is hard to compare novel approaches with the previous ones in an objective manner.

In this paper, we propose a novel large scale database consisting of 48,000 samples, generated automatically while trying to replicate a realistic automotive scenario with one interference source. We generate every sample using specific randomly selected values for the set of parameters listed in Table~\ref{tab1}. While the values corresponding to some parameters are selected using a fixed step between the minimum and the maximum values specified in Table~\ref{tab1}, the values corresponding to the other parameters are randomly selected using an uniform distribution between the minimum and the maximum values. More precisely, we use linear variation with a fixed step for the signal to noise ratio (SNR), the signal to interference ratio (SIR) and the interference slope parameters. The number of targets as well as the distance, the amplitude and the phase of each target are random variables that follow an uniform distribution. The amplitude of each target is proportional with the power expected from that particular target. We added a random phase to each target to obtain more realistic radar signals. 

Concerning the simulated radar sensor, we considered a fixed set of parameters such as bandwidth, sweep time, sampling frequency and central frequency. The exact values used for these parameters are listed in Table~\ref{tab2}.

Since we can control all factors during sample generation, we can produce an exact copy of each signal without interference. First of all, the clean copy can be used as ground-truth label when training a machine learning model. Second of all, it provides the means to conduct an objective assessment of the performance, by comparing the output predicted by the model with the corresponding ground-truth (expected) output. Consequently, a data sample is composed of:
\begin{itemize}
\item a time domain signal without interference; 
\item a time domain signal with interference;
\item a label vector with complex amplitude values in target locations.
\end{itemize}

We randomly split our data samples into a training set of 40,000 samples and a test set of 8,000 samples. This will allow future works to directly compare novel results with previous results, without having to re-implement preceding methods in order to reproduce the corresponding results. Our data set is freely available for download at: http://github.com/ristea/arim.

\section{Experiments}

Since the database consists of different radar signals (with and without interference) referring to different range profiles, in our experiments, the interference mitigation is performed individually on each range profile.

\noindent
{\bf Evaluation metrics.}
Usually, the goal in radar signal processing is to maximize the detection performance. Thus, a rather intuitive measure is the area under the receiver operating characteristics curve (AUC), which describes the ability to disentangle targets from noise at various thresholds. The target detection threshold grows iteratively from the lowest value to the largest value in range profile, modifying the probability of false alarms. Another performance indicator is the mean absolute error (MAE) in decibels (dB) between the range profile amplitude of targets computed from label signals and the amplitude of targets from predicted signals. In our evaluation, we employed the AUC, the MAE and the mean SNR improvement ($\overline{\Delta \mbox{SNR}}$), which is computed for the target with the highest amplitude in a signal as the difference between SNR after and before interference mitigation.

\noindent
{\bf Baseline.}
A very common approach to eliminate the interference is to replace amplitudes higher than a specific threshold with zero (e.g. \cite{mosarim2010d15,laghezza2019nxp}). This method is denoted as \emph{zeroing}.

\noindent
{\bf Hyperparameter tuning.}
We tune the hyperparameters of our FCN models on a validation set. We kept 20\% of the training set (8,000 samples) for validation. We used the same hyperparameters for both architectures, in order to minimize the chance of overfitting in hyperparameter space. We trained the models for 100 epochs with a mini-batch size of 10 samples. We set the learning rate to $10^{-5}$ and we used a weight decay of $10^{-5}$. These parameters are obtained using grid search. In a similar fashion, we employed grid search on the validation set to find the threshold parameter for our baseline, the zeroing method.

\begin{table}[t]
\centering
\noindent
\caption{Validation and test results provided by our shallow and deep FCN models versus an oracle based on true labels and a baseline based on zeroing. A higher $\overline{\Delta \mbox{SNR}}$ value is better, a higher AUC value is better and a lower MAE value is better. The best results (excluding the oracle) are highlighted in bold.\vspace{0.1cm}}
\setlength\tabcolsep{4.5pt}
\begin{tabular}{|l|ccc|ccc|}
\hline
 & \multicolumn{3}{|c|}{{Validation set}} & \multicolumn{3}{|c|}{{Test set}} \\ 
\hline
{{Method}} & {$\overline{\Delta \mbox{SNR}}$} & {AUC} & {MAE}& {$\overline{\Delta \mbox{SNR}}$} & {AUC} & {MAE} \\
 &  & & {(dB)} & & & {(dB)} \\ 
\hline
\hline
Oracle (true labels) & 12.92 & 0.978 & 0 & 13.08 & 0.978 & 0 \\ 
\hline
{Zeroing} & 5.27 & 0.951 & 1.26 & 5.44 & 0.951 & 1.27 \\ 
{Shallow FCN} & 10.34 & 0.965 & 2.20 & 10.49 & 0.965 & 2.21 \\ 
{Deep FCN} & \textbf{12.90} & \textbf{0.972} & \textbf{1.21} & \textbf{13.06} & \textbf{0.972} & \textbf{1.22} \\ 
\hline
\end{tabular}
\label{tab_results}
\vspace{-0.1cm}
\end{table}

\noindent
{\bf Results.}
We compared our FCN models with a zeroing benchmark and an oracle based on the ground-truth labels. We present the corresponding results in Table~\ref{tab_results}. We included the oracle in order to show the achievable upper bound score for each metric, e.g. the maximum AUC score on both validation and test is 0.978. Comparing our shallow FCN model with the zeroing baseline, we observe that our network attains better mean SNR improvement and AUC scores, but it seems to underperform according to the MAE measure. Our deep FCN provides superior results for all three metrics, surpassing both the zeroing baseline and the shallow FCN. In terms of $\overline{\Delta \mbox{SNR}}$ and AUC, our deep FCN seems to attain performance levels quite close to the oracle, e.g. the AUC of our deep FCN is 0.006 under the AUC of the oracle on the test set.


\begin{figure}[t!]
\centering
 \includegraphics[width=1.0\linewidth]{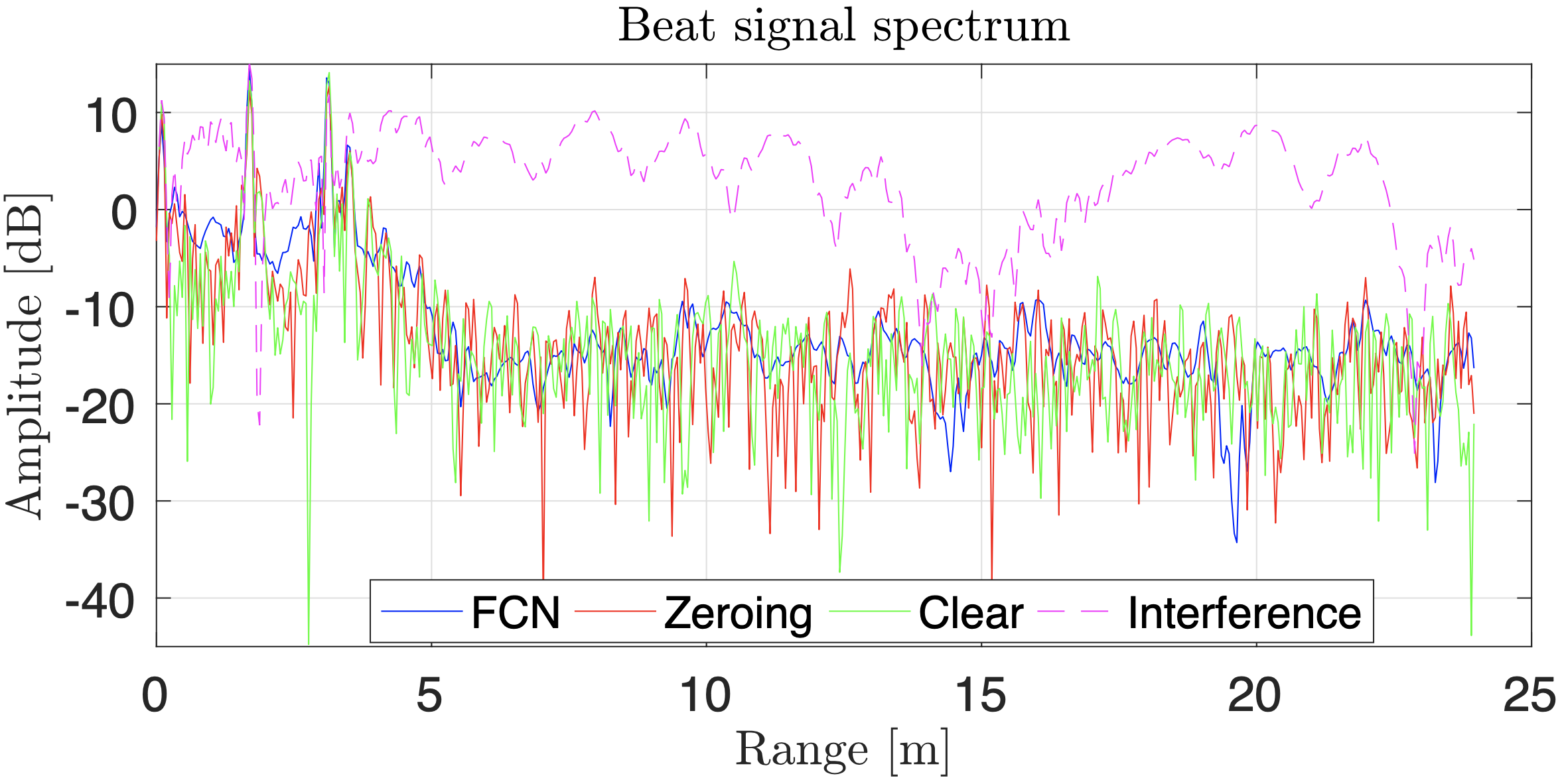}
  \vspace{-0.5cm}
  \caption{Results for radar interference mitigation with our best FCN model. For comparison, we also added the ground-truth signal, the signal with interference and the signal with mitigated interference by the zeroing method.}
  \label{fig_3}
  \vspace{-0.3cm}
\end{figure}

In addition to the results on our data set, we assessed the generalization capacity of our deep FCN on real data, by testing it on samples provided by the NXP company, which were captured with the NXP TEF810X 77 GHz radar transceiver, from real-world scenarios. In Figure~\ref{fig_3}, we show an example of interference mitigation performed by our model on a real radar signal, producing an output very similar to the reference signal.


\section{Conclusion}
In this paper, we introduced two fully convolutional networks for automotive radar interference mitigation and a large scale database of radar signals simulated in realistic settings. We compared our FCN models with the zeroing baseline in a comprehensive experiment, showing that our deeper FCN provides superior results. We also released our novel data set to allow objective comparison in future work. To our knowledge, we are the first to establish a benchmark data set for automotive radar interference mitigation. In future work, we aim to analyze the scenario with multiple interference sources.


\bibliographystyle{ieeetr}
\bibliography{report}

\end{document}